\documentclass[twocolumn,usenatbib]{aastex62}

\usepackage{latexsym}
\usepackage{color}
\usepackage{amsmath}
\usepackage{amssymb}
\usepackage{graphicx}
\usepackage{aas_macros}
\usepackage{comment}

\graphicspath{{./}{figures/}}

\submitjournal{ApJL}

\shorttitle{Pop~III origin of GW190521}
\shortauthors{Liu, B. \& Bromm, V.}

\begin{document}

\title{The Population~III origin of GW190521}

\correspondingauthor{Boyuan Liu}
\email{boyuan@utexas.edu}

\author[0000-0002-4966-7450]{Boyuan Liu}
\affiliation{Department of Astronomy, 
University of Texas at Austin,
TX 78712, USA}

\author{Volker Bromm}
\affiliation{Department of Astronomy,
University of Texas at Austin,
TX 78712, USA} 

\begin{abstract}
We explore the possibility that the recently detected black hole binary (BHB) merger event GW190521 originates from the first generation of massive, metal-free, so-called Population~III (Pop~III), stars. Based on improved binary statistics derived from N-body simulations of Pop~III star clusters, we calculate the merger rate densities of Pop~III BHBs similar to GW190521, in two evolution channels: classical binary stellar evolution and dynamical hardening in high-redshift nuclear star clusters. Both channels can explain the observed rate density. But the latter is favoured by better agreement with observation and less restrictions on uncertain parameters. Our analysis also indicates that given the distinct features of the two channels (with merger rates peaked at $z\lesssim2$ and $z\sim 10$, respectively), future observation of BHB mergers similar to GW190521 with third-generation gravitational wave detectors will greatly improve our knowledge of the evolution of Pop~III BHBs, especially for their dynamics during cosmic structure formation.
\end{abstract}

\keywords{early universe --- dark ages, reionization, first stars --- stars: Population III --- gravitational waves}

\section{Introduction}
\label{s1}
The black hole binary (BHB) merger event GW190521 detected at a redshift of $0.82_{-0.34}^{+0.28}$ during the third observing run (O3) of LIGO/Virgo has unusual BH masses $85_{-14}^{+21}\ \rm M_{\odot}$ and $66_{-18}^{+17}$ \citep{abbott2020gw190521,abbott2020properties}, right within the mass gap $\sim 55-130\ \rm M_{\odot}$ predicted by standard pulsational pair-instability supernova (PPISN) models (e.g., \citealt{heger2003massive,belczynski2016effect, woosley2017pulsational,marchant2019pulsational}). After its announcement, many studies explored the properties and origin of this unique event, including its statistical interpretation and implication \citep{fishbach2020don,wang2020gw190521}, highly eccentric orbit \citep{gayathri2020gw190521}, uncertainty of the mass gap \citep{costa2020formation}, repeated BH mergers and stellar collisions in dense star clusters (e.g., \citealt{fragione2020origin,romero2020gw190521,kremer2020populating,renzo2020stellar}), mergers of ultra dwarf galaxies \citep{palmese2020}, accretion in dense molecular clouds \citep{rice2020growth}, and primordial BHs \citep{de2020gw190521}. 

\begin{figure}[htbp]
    \centering
    \includegraphics[width=1\columnwidth]{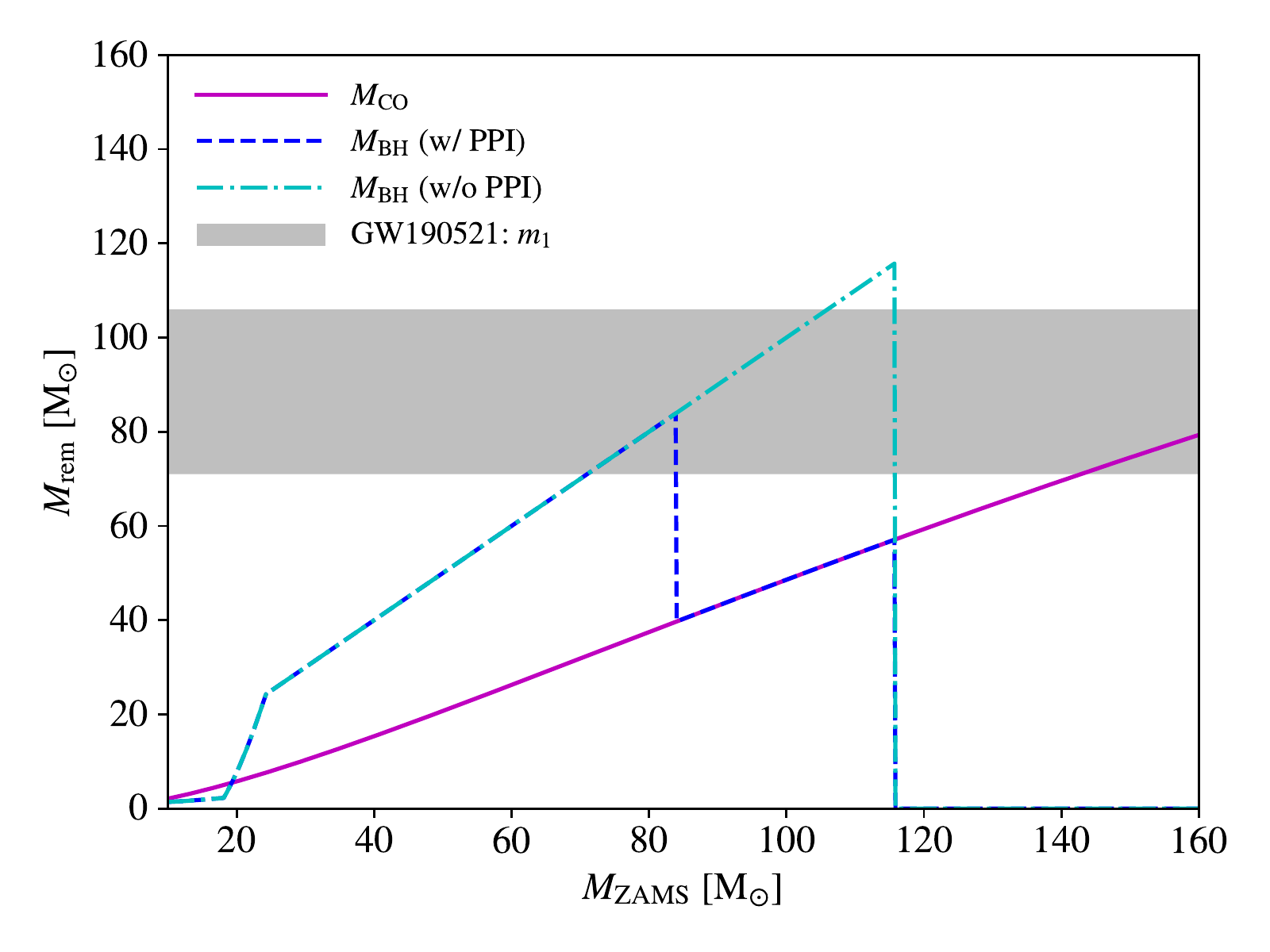}
    \caption{Pop~III remnant mass with (dashed) and without (dashed dotted) pulsational pair-instability (PPI) effect, and CO core mass (solid), as functions of ZAMS mass, for the $Z=10^{-6}\ \rm Z_{\odot}$ model in \citet{tanikawa2020fitting}. The shaded region shows the 90\% likelihood interval for the primary mass in GW190521 \citep{abbott2020gw190521}.}
    \label{f1}
\end{figure}

Particularly, as discussed in \citet{farrell2020gw190521,tanikawa2020population}, the first generation of stars in the universe, so-called Population~III (Pop~III) stars with zero or very low metallicities, are promising progenitors of the BHs found in GW190521, because Pop~III stars, with small sizes and little mass loss, are likely to retain most of their hydrogen envelopes until the pre-SN stage, avoid the PPISN regime and form BHs in the mass gap (reaching up to $\sim 85\ \rm M_{\odot}$) with fallback (see Fig.~\ref{f1}). 
Actually, \citet{kinugawa2020formation} show that classical binary stellar evolution models for isolated binaries can explain the observed merger rate density of events like GW190521, 
under certain assumptions \citep{kinugawa2020}. 

However, the uncertainties in binary stellar evolution models are significant, for (initial) binary statistics, common envelope (CE) parameters and SN kicks, leading to up to two orders of magnitude discrepancy in the BHB merger rate (e.g., \citealt{kinugawa2014possible,belczynski2017likelihood}). It remains unknown whether a Pop~III star can keep its hydrogen envelope to reach $\sim 85\ \rm M_{\odot}$ and meanwhile experience close binary interactions 
\citep{farrell2020gw190521}. 

Furthermore, as Pop~III stars are mostly formed at high redshifts (peaked at $z\sim 10$; e.g., \citealt{johnson2013first,xu2016late,popIIIterm}), their remnants will be affected by the entire process of cosmic structure formation, and it is necessary to consider the effect of environment for the evolution of Pop~III BHs/BHBs. For instance, it is shown that Pop~III BHs with initial masses below the mass gap can grow to $\sim 85\ \rm M_{\odot}$ via rapid gas accretion, and thus form BHBs similar to GW190521, in centers of atomic-cooling halos \citep{safarzadeh2020}, protoglobular clusters with early ($\lesssim10$~Myr) growth before gas depletion \citep{roupas2019generation}, and nuclear star clusters with wind-fed supra-exponential accretion \citep{natarajan2020new}. Moreover, interactions with surrounding stars can allow initially wide binaries to merge within a Hubble time, without undergoing close binary interactions. In light of this, we explore the possible Pop~III origin of GW190521, focusing on an alternative channel of dynamical hardening in dense star clusters. 


\section{Pop~III BH binary models}
\label{s2}
Our analysis is based on improved binary statistics derived from N-body simulations of Pop~III star clusters, employing a novel physically-motivated model for the initial cluster configuration, described in \citet{boyuan2020c}. The main advantage of our approach is the consideration of the (nearly) self-similar nature of disc evolution and fragment properties during hierarchical fragmentation, inferred from small-scale hydrodynamic simulations of Pop~III protostellar systems (e.g., \citealt{hirano2017formation,susa2019merge,sugimura2020birth}). Previous studies (e.g., \citealt{kinugawa2014possible,belczynski2017likelihood,kinugawa2020}) apply the binary parameters of present-day stars or Pop~III protostars (of only a few $\rm M_{\odot}$) to much more massive newly formed Pop~III systems (after gas removal by feedback), resulting in smaller clusters than expected from angular-momentum conservation, and thus likely overestimate the fraction of stars in close binaries. 

We adopt the fiducial (FD) model, \texttt{tf1e2ta1e5a1m1}, fully described in \citet{boyuan2020c}, in which Pop~III BHBs with primary masses $m_{1}\sim 71-106\ \rm M_{\odot}$ and total masses $m\equiv m_{1}+m_{2}\sim 133-179\ \rm M_{\odot}$ are identified as candidate progenitors for GW190521, according to the 90\% probability intervals for these parameters inferred from observation. We use the fitting formulae for Pop~III stellar evolution at a metallicity of $Z=10^{-6}\ \rm Z_{\odot}$ from \citet{tanikawa2020fitting} to map zero-age main sequence (ZAMS) to BH mass, including two extreme cases (Fig.~\ref{f1}). In the optimistic case, there is no pulsational pair-instability (PPI) effect, such that $M_{\rm BH}\sim M_{\rm ZAMS}$. In the pessimistic case, a star with PPI ($M_{\rm ZAMS}\sim 85-115\ \rm M_{\odot}$) will lose its envelope, such that $M_{\rm BH}=M_{\rm CO}\sim 40-55\ \rm M_{\odot}$.
We consider two channels for Pop~III BHB mergers: classical binary stellar evolution for close binaries with CE evolution, and dynamical hardening (DH) for wide binaries in dense star clusters.

\subsection{Classical binary stellar evolution}
\label{s2.1}
For simplicity, we here estimate the efficiency of producing BHB mergers like GW190521 from Pop~III stars via classical binary stellar evolution without detailed modelling of the mass transfer and CE evolution, as well as SN kicks. Instead, we identify all binaries with Roche lobe overflow of either one of the two stars that is massive enough ($M_{\rm ZAMS}> 50~\rm M_{\odot}$, \citealt{kinugawa2020}) to have a convective envelope. 
We assume that these binaries will undergo CE evolution due to unstable mass transfer, and merge within a Hubble time, $t_{\rm H}\simeq 13.7$~Gyr, following a power-law delay time distribution $\mathcal{P}(t_{\rm GW})$ with a slope of -1 in the range of 3 to $10^{4}$~Myr, based on previous studies (e.g., \citealt{belczynski2017likelihood})\footnote{The results are not sensitive to the detailed shape of $\mathcal{P}(t_{\rm GW})$ as long as it is dominated by mergers with short delay times (a few Myr), which is a common outcome in classical binary stellar evolution models (e.g., \citealt{belczynski2017likelihood,kinugawa2020,tanikawa2020merger}).}. 

Because close binaries with $a\lesssim 10$~AU in the FD model are rare, the efficiency of Pop~III BHB mergers\footnote{The efficiency of BHB mergers for a stellar population is defined as the average number of BHBs that can merge within a Hubble time per unit stellar mass.} similar to GW190521 via CE evolution with PPI is low, $f_{\rm BHB}=3\times 10^{-6}\ \rm M_{\odot}^{-1}$. Considering other models in \citet{boyuan2020c} with much (a factor of $\gtrsim 15$) smaller cluster sizes and more close binaries, we obtain an extreme upper limit of $f_{\rm BHB}\sim 10^{-4}\ \rm M_{\odot}^{-1}$ in the optimistic case. Note that we do not include SN kicks for the primary, which can enhance the chance of CE evolution by shrinking the binary orbit, especially important for BHBs in the PPISN mass gap (see fig.~33 of \citealt{tanikawa2020merger}). Therefore, we may have underestimated the efficiency of BHB mergers driven by CE evolution.

\subsection{Dynamical hardening in dense star clusters}
\label{s2.2}
We next consider how BHB evolution depends on environment, which can drive initially wide binaries to merge without a CE phase that is only available for initially close binaries. Actually, the dynamics of BHB coalescence via environmental effects is very complex, involving various astrophysical aspects such as a clumpy interstellar medium, dark matter distribution, nuclear star clusters (NSCs), supermassive BHs and AGN discs (e.g., \citealt{rovskar2015orbital,antonini2016merging,tamfal2018formation,leigh2018rate,choksi2019star,ogiya2019,zhang2019gravitational,secunda2019orbital}). For simplicity, we focus on one mechanism particularly important for massive Pop~III BHBs: the dynamical hardening (DH) by 3-body interactions with surrounding (low-mass) stars in dense star clusters \citep{antonini2016merging,leigh2018rate}. Below we estimate the delay time for a BHB merger with DH.

Following \citet{boyuan2020}, the evolution of the semi-major axis $a$ for a BHB driven by 3-body interactions with surrounding stars and gravitational wave (GW) emission, can be written as \citep{sesana2015scattering} 
\begin{align}
\frac{da}{dt}=\left.\frac{da}{dt}\right\vert_{\mathrm{3B}}+\left.\frac{da}{dt}\right\vert_{\mathrm{GW}}=-Aa^{2}-\frac{B}{a^{3}}\ .\label{e1}
\end{align}
The first term represents 3-body interactions, the second the energy and angular momentum loss via GWs, and
\begin{align}
\begin{split}
A&=\frac{GH\rho_{\star}}{\sigma_{\star}}\ ,\\
B&=\beta F(e)\ ,\quad \beta=\frac{64G^{3}m_{1}m_{2}m}{5 c^{2}}\ .
\end{split}\label{e2}
\end{align} 
Here, $m_{1}$ and $m_{2}$ are the masses of the primary and secondary, $m\equiv m_{1}+m_{2}$, $\sigma_{\star}$ and $\rho_{\star}$ are the velocity dispersion and stellar density around the BHB, $F(e)=(1-e^{2})^{-7/2}[1+(73/24)e^{2}+(37/96)e^{4}]$ given the eccentricity $e$, and $H\sim 15-20$ is a dimensionless parameter, which we set to $17.5$ for simplicity.

The binary system first undergoes a phase dominated by DH with a characteristic semimajor axis, estimated by imposing $(da/dt)|_{\mathrm{3B}}=(da/dt)|_{\mathrm{GW}}$: $a_{\star/\mathrm{GW}}=\left(B/A\right)^{1/5}$. 
We assume a constant eccentricity in the DH-dominated phase for simplicity. The duration of this stage for $a$ to reach $a_{\star/\mathrm{GW}}$, the DH timescale, can be estimated with
\begin{align}
t_{\mathrm{DH}}= \frac{1}{Aa_{\star/\mathrm{GW}}}-\frac{1}{Aa_{0}}= \left(\frac{1}{A^{4}B}\right)^{1/5}-\frac{1}{Aa_{0}}\ ,\label{e3}
\end{align}
where $a_{0}$ is the initial semi-major axis. We set $t_{\rm DH}=0$, $a_{\star/\rm GW}=a_{0}$ for $a_{0}<(B/A)^{1/5}$.

After the DH-dominated phase ($a\lesssim a_{\star/\rm GW}$), the evolution is dominated by GW emission. The time spent in GW-driven inspiral before the final coalescence can be estimated with
\begin{align}
\begin{split}
&t_{\mathrm{col}}=\frac{12}{19}\frac{c_{0}^{4}}{\beta}\int_{0}^{e}dx\frac{x^{29/19}[1+(121/304)x^{2}]^{\frac{1181}{2299}}}{(1-x^{2})^{3/2}}\ ,\\
&c_{0}=\frac{a_{\star/\mathrm{GW}}(1-e^{2})}{e^{12/19}}\left[1+\frac{121}{304}e^{2}\right]^{-870/2299}\ ,
\end{split}\label{e4}
\end{align}
The total time of inspiral 
in the dense star cluster is then given by $t_{\rm SC}=t_{\rm DH}+t_{\rm col}$. Other than the binary parameters ($m_{1}$, $m_{2}$, $e$ and $a_{0}$), the key parameters that determine $t_{\rm SC}$ are $\rho_{\star}$ and $\sigma_{\star}$. Besides, only \textit{hard} binaries will be further hardened (instead of destroyed) by 3-body interactions, which must satisfy $Gm_{1}m_{2}(1-e)/[a(1+e)]>m_{\star}\sigma_{\star}^{2}$ in order to survive disruptions close to the apocenter, where $m_{\star}=1\ \rm M_{\odot}$ is the typical mass of surrounding stars.

\subsection{Pop~III BHBs in high-$z$ NSC hosts}
\label{s2.3}

To apply the DH machinery to Pop~III BHBs, we need to know the properties of host clusters ($\rho_{\star}$ and $\sigma_{\star}$), initial binary parameters and dynamics of Pop~III BHs/BHBs falling into dense star clusters. Here we consider Pop~III BHBs in nuclear star clusters (NSCs), which occupy the innermost regions of most galaxies (\citealt{neumayer2020nuclear}), and are also common in high-$z$ atomic-cooling halos with $M_{\rm h}\gtrsim 10^{8}\ \rm M_{\odot}$ (e.g., \citealt{devecchi2009formation,devecchi2010high,devecchi2012high}). 


For NSC properties, we explore three cases with $\sigma_{\star}\simeq 10,\ 33\text{ and }100\ \rm km\ s^{-1}$, $\rho_{\star}=10^{4},\ 7\times 10^{5}\text{ and }5\times 10^{7}\ \rm M_{\odot}\ pc^{-3}$, corresponding to NSCs in halos with masses $M_{\rm h}\sim 10^{8},\ 10^{10}\text{ and } 10^{12}\ \rm M_{\odot}$, 
respectively, which dominate the halo populations at $z_{\rm crit}\sim 9,\ 5.5$ and 4 for $2\sigma$ peaks. These three models are meant to cover most ($\gtrsim90\%$) of the region occupied by observed NSCs with masses $M_{\rm NSC}\lesssim 10^{7}\ \rm M_{\odot}$ for $M_{\rm h}\lesssim 10^{12}\ \rm M_{\odot}$ (see fig.~7 of \citealt{neumayer2020nuclear}) in the parameter space. In reality, the host NSC properties exhibit scatters and also evolve with redshift. 
The realistic situation will be a mixture of our models. To populate the $\rho_{\star}$-$\sigma_{\star}$ space with observed NSCs (see Fig.~\ref{fa1} in Appendix~\ref{a1}), we estimate $\sigma_{\star}$ with $\sqrt{(1/2)GM_{\rm NSC}/r_{\rm eff}}$, given the half-mass(light) radius $r_{\rm eff}$. We set $\rho_{\star}$ to the stellar density within a characteristic radius $r_{\rm BH}$ where the enclosed stellar mass is twice the mass of the infalling BHB ($m\sim 140\ \rm M_{\odot}$), assuming that the stars follow the Dehnen profile \citep{dehnen1993family} with an inner slope of $\gamma=1$ (see Sec.~4.1.2 in \citealt{boyuan2020})\footnote{$\gamma=1$ is a conservative choice, which is also consistent with the trend $r_{\rm eff}\propto M_{\rm NSC}^{1/2}$ \citep{neumayer2020nuclear}.}. Here $r_{\rm BH}$ describes how deep the BHB can sink into the cluster by mass segregation. 

We substitute the candidate BHB progenitors of GW190521 
from the FD model into the DH scheme (Equ.~\ref{e1}-\ref{e4}), 
in which only \textit{hard} binaries without CE evolution are considered. In this way, we derive the NSC inspiral time distributions $p(t_{\rm SC})$ for the three NSC models. The results with PPI can be well fitted by $p(t_{\rm SC})\propto t_{\rm SC}^{\alpha}$ with $\alpha=2$, 4 and 8, in the ranges of $2.5-13$~Gyr, $0.44-1.1$~Gyr and $69-90$~Myr, for $M_{\rm h}\sim 10^{8},\ 10^{10}\text{ and } 10^{12}\ \rm M_{\odot}$, respectively (see Fig.~\ref{fa3} in Appendix~\ref{a1}). The corresponding efficiencies of mergers are $f_{\rm BHB,0}=1.3\times 10^{-4}$, $7.6\times 10^{-5}$ and $7.8\times 10^{-6}\ \rm M_{\odot}$. Turning off PPI will enhance $f_{\rm BHB,0}$ by a factor of $\sim 5$, with $p(t_{\rm SC})$ almost unchanged. The final efficiency is given by $f_{\rm BHB}=f_{\rm NSC}f_{\rm BHB,0}$, where $f_{\rm NSC}\sim 0.2-0.8$ \citep{neumayer2020nuclear} is the fraction of galaxies that host NSCs.

Now, taking into account the BH/BHB dynamics in host halos during cosmic structure formation, the final delay time $t_{\rm GW}$ can be written as $t_{\rm GW}=t_{\rm SC}+t_{\rm IF}+t_{\rm DF}$, where $t_{\rm IF}$ and $t_{\rm DF}$ are the timescales for the BHB to fall into the \textit{final} host halo and sink into the NSC via dynamical friction of gas. In general, Pop~III star formation peaks at $z\sim 10$ in low-mass halos ($M_{\rm h}\sim 10^{7}-10^{8}\ \rm M_{\odot}$) at $2\sigma$ peaks, and the host halos of Pop~III BHs will then grow or merge into more massive halos, in which the Pop~III BHs can further sink into NSCs. The halo growth/merger process is captured by the timescale $t_{\rm IF}$. Given the mass $M_{\rm h}$ of the final host halo, $t_{\rm IF}$ can be estimated with $t(z=z_{\rm crit})-t(z=10)$, where $t$ is the age of the universe as a function of redshift, and $z_{\rm crit}$ is the redshift at which $M_{\rm h}$ matches the critical mass at $2\sigma$ peaks. We have $t_{\rm IF}\simeq 0.07$, 0.56 and 1.06~Gyr, for $M_{\rm h}\sim 10^{8},\ 10^{10}\text{ and } 10^{12}\ \rm M_{\odot}$, respectively. 

The (gas) dynamical friction timescale $t_{\rm DF}$ depends on the (initial) distance $r$ of the BHB to the galaxy center and the gas density profile. We adopt the $t_{\rm DF}$ formula in \citet[][see their equ.~18 and table~1]{boco2020growth} with fixed Coulomb logarithm and assuming circular orbits for the BHBs. This formula can be applied to different density profiles, parameterized by the gas mass $M_{\rm gas}$ and half-mass radius $R_{e}$. For the $M_{\rm h}=10^{8}\ \rm M_{\odot}$ model at $z=9$, which is a typical high-$z$ atomic cooling halo, the singular isothermal sphere (SIS) is a good approximation to the density profiles found in simulations at $r\gtrsim 1$~pc (see equ.~2 in \citealt{safarzadeh2020}). We consider the entire halo\footnote{The $t_{\rm DF}$ formula in \citealt{boco2020growth} does not include the effect of dark matter, which is valid for the inner regions ($\lesssim 1$~kpc) of massive halos ($M_{\rm h}\gtrsim 10^{10}\ \rm M_{\odot}$). However, in our case with $M_{\rm h}=10^{8}\ \rm M_{\odot}$ at $z=9$, one has to consider dark matter even for $r\lesssim 1$~kpc, as the virial radius is $R_{\rm vir}\simeq 1.44$~kpc. Therefore, we boost the initial specific angular momentum of the BHB by a factor of $(\Omega_{\rm m}/\Omega_{\rm b})^{1/2}$ to include the gravity of dark matter. The dynamical friction timescale obtained in this way is consistent with that in \citet{safarzadeh2020} within a factor of 2.} with $M_{\rm gas}\sim M_{\rm h}\Omega_{\rm b}/\Omega_{\rm m}=1.55\times 10^{7}\ \rm M_{\odot}$ and $R_{e}=0.5 R_{\rm vir}=0.72$~kpc, which gives $t_{\rm DF}\sim 140\ \mathrm{Myr}\times (r/10\ \mathrm{pc})^{2}$. For the $M_{\rm h}=10^{12}\ \rm M_{\odot}$ model at $z=4$, as a typical progenitor for local early-type galaxies, we follow \citet{boco2020growth} to focus on the gas-dominated inner region with $R_{e}=1$~kpc and $M_{\rm gas}\sim 2(R_{e}/R_{\rm vir})^{0.3}M_{\rm h}\Omega_{\rm b}/\Omega_{\rm m}= 9\times 10^{10}\ \rm M_{\odot}$, under a Sersic profile with index $n=1.5$, such that $t_{\rm DF}\sim 150\ \mathrm{Myr}\times (r/10\ \mathrm{pc})^{2.7}$. Finally, for the intermediate case with $M_{\rm h}=10^{10}\ \rm M_{\odot}$ at $z=5.5$, we adopt the Hernquist profile, which is between the aforementioned SIS and Sersic models in terms of compactness. Again, we consider the gas-dominated inner region with $R_{e}=1$~kpc and $M_{\rm gas}\sim 2(R_{e}/R_{\rm vir})^{0.5}M_{\rm h}\Omega_{\rm b}/\Omega_{\rm m}=9.7\times 10^{8}\ \rm M_{\odot}$, leading to $t_{\rm DF}\sim 40\ \mathrm{Myr}\times (r/10\ \mathrm{pc})^{2.5}$. Although our modelling of gas distributions is highly idealized, it captures the trend that gas is more concentrated due to stronger cooling (with higher metallicities) in more massive halos at lower redshifts. 

\begin{figure}[htbp]
    \centering
    \includegraphics[width=1\columnwidth]{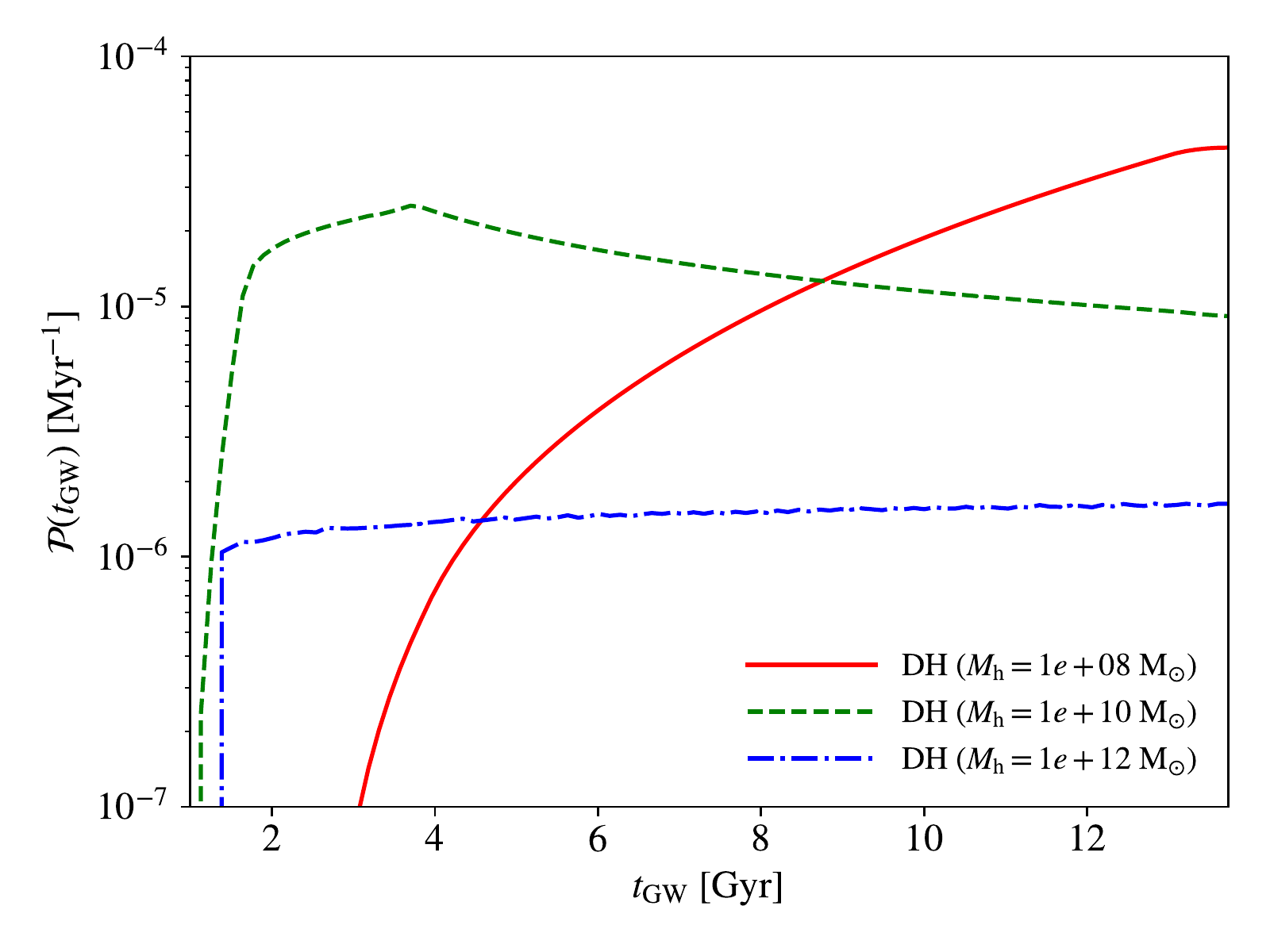}
    \caption{Delay time distributions of Pop~III BHB mergers similar to GW190521, driven by DH in NSCs hosted by halos with masses $M_{\rm h}\sim 10^{8},\ 10^{10}\text{ and } 10^{12}\ \rm M_{\odot}$ (typical at $z=9$, 5.5 and 4 for $2\sigma$ peaks), denoted by the solid, dashed and dashed-dotted curves, respectively. }
    \label{f3}
\end{figure}

Finally, we consider the distribution of Pop~III BHs/BHBs in high-$z$ galaxies. We derive the fraction of Pop~III BHs $f_{\rm BH}(<r)$ enclosed within $r$ around galaxy centers in atomic-cooling halos ($M_{\rm h}\gtrsim 10^{8}\ \rm M_{\odot}$), from the cosmological simulation in \citet[][see Fig.~\ref{fa2} in Appendix~\ref{a1}]{boyuan2020}, where the dynamical friction of gas is unresolved. Then the (normalized) distribution of $t_{\rm GW}$ takes the form $\mathcal{P}(t_{\rm GW}|M_{\rm h})=\int_{r_{\min}}^{r_{\max}} dr \left[p(t_{\rm GW}-t_{\rm IF}(M_{\rm h})-t_{\rm DF}(r|M_{\rm h}))df_{\rm BH}(<r)/dr\right]$, where we adopt $r_{\min}=10$~pc and $r_{\max}=10^{3}$~pc, since $f_{\rm BH}(<r)$ at $r<10$~pc is negligible ($\lesssim 10^{-3}$), and $t_{\rm DF}(r|M_{\rm h})>t_{\rm H}$ for $r>10^{3}$~pc. The resulting distributions are shown in Fig.~\ref{f3}.


\section{Resulting merger rate}
\label{s4}
\begin{figure}[htbp]
\centering
\vspace{5pt}
\includegraphics[width=1\columnwidth]{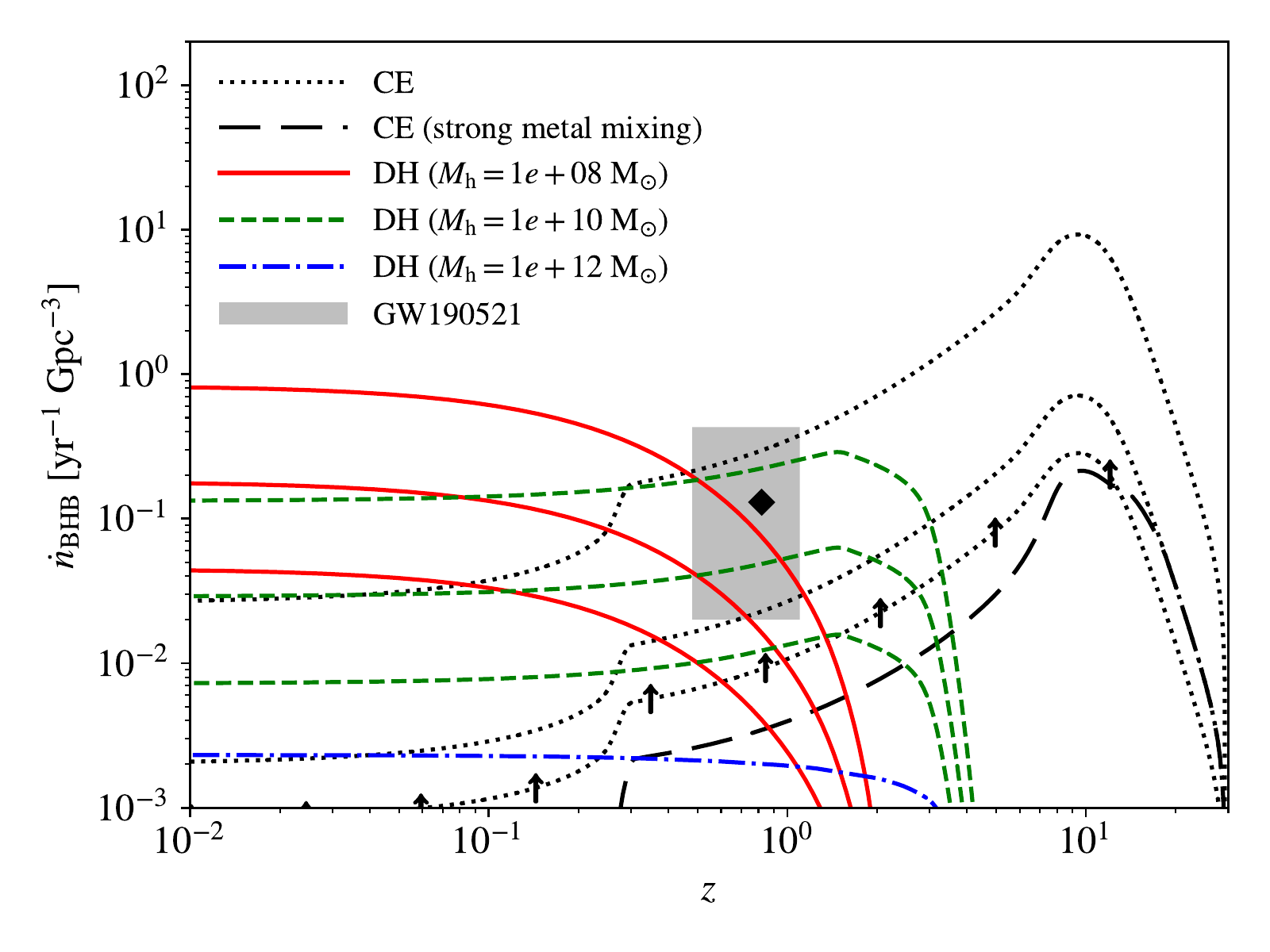}
\caption{Co-moving rate density of Pop~III BHB mergers similar to GW190521 as a function of redshift. The rates inferred from GW190521 are shown with the shaded region (for 90\% confidence intervals) and diamond symbol (for best-fit values, \citealt{abbott2020gw190521}). 
Predictions from the three NSC DH models (Sec.~\ref{s2.3}) are shown with solid, dashed and dashed-dotted curves, for $M_{\rm h}\sim 10^{8},\ 10^{10}\text{ and } 10^{12}\ \rm M_{\odot}$ (typical at $z=9$, 5.5 and 4 for $2\sigma$ peaks), respectively. Each model has three curves, where the middle and bottom ones correspond to the upper and lower bounds for the occupation fraction of NSCs $f_{\rm NSC}\sim 0.2-0.8$, with PPI included, while the top one is the upper limit with optimal $f_{\rm NSC}$ and no PPI. The results for CE evolution (Sec.~\ref{s2.1}) are shown with dotted and long-dashed curves. The middle and lower dotted curves (not shown for the $M_{\rm h}=10^{12}\ \rm M_{\odot}$ case) correspond to the optimistic and pessimistic cases (without and with PPI effect) from the FD model in \citet{boyuan2020c}. While the upper dotted curve corresponds to the extreme upper bound with no PPI effect and smaller cluster sizes. For the long-dashed curve, we adopt the Pop~III SFRD with strong metal mixing and the optimistic $f_{\rm BHB}$ from the FD model. }
\label{master}
\end{figure}

Once the efficiency $f_{\rm BHB}$ and delay time distribution $\mathcal{P}(t_{\rm GW})$ of Pop~III BHBs similar to GW190521 are known from a given evolution model, the merger rate density can be calculated by convolving with the input Pop~III star formation rate density (SFRD) $\dot{\rho}_{\star,\rm PopIII}$:
\begin{align}
    \dot{n}_{\rm BHB}(t)=\int_{0}^{t-t_{i}}f_{\rm BHB}\mathcal{P}(t')\dot{\rho}_{\star,\rm PopIII}(t-t')dt'\ ,
\end{align}
where 
$t$ is the age of the universe, and $t_{i}\sim 100$~Myr (corresponding to $z\sim 30$) marks the onset of Pop~III star formation. We apply the optimistic Pop~III SFRD model in \citet[][see the thick solid line in their fig.~13]{popIIIterm} to the Pop~III BHB evolution models discussed above. For the optimal CE evolution model, we further consider a pessimistic SFRD case under strong metal mixing (see the long-dashed curve in fig.~13 of \citealt{popIIIterm}), where late-time ($z\lesssim 6$) Pop~III star formation is suppressed. 
The resulting evolution of $\dot{n}_{\rm BHB}$ in a variety of models is shown in Fig.~\ref{master}. 

It turns out that the CE model can only marginally reproduce the merger rate inferred from GW190521 with the optimistic efficiency from the FD model and Pop~III SFRD, or with significant enhancement by SN kicks. However, if strong metal mixing is present, the merger rate from the CE channel is lower than the observed rate by a factor of $\sim 5-100$, even with the optimistic efficiency from the FD model. If Pop~III clusters are much (a factor of $\gtrsim 15$) smaller than in the FD model, a better agreement can be achieved. But such cases seem unlikely to occur \citep{boyuan2020c}. The NSC DH models for $M_{\rm h}=10^{8}\text{ and }10^{10}\ \rm M_{\odot}$ at $z\gtrsim 5.5$ agree well with the observational constraint. While the $M_{\rm h}=10^{12}\ \rm M_{\odot}$ model fails to drive enough Pop~III BHBs into NSCs, and thus, underpredicts the meger rate even in the optimistic case. 

\section{Summary and Conclusions}
\label{s5}
We explore the possible Pop~III origin of the recently reported BHB merger event GW190521 with BH masses in the pulsational pair-instability supernova (PPISN) mass gap \citep{abbott2020gw190521}. In particular, we consider the channel of dynamical hardening (DH) in high-$z$ nuclear star clusters (NSCs) for initially wide Pop~III BHBs to merge within a Hubble time, in difference from the classical binary stellar evolution channel via common envelope (CE) evolution of close binaries \citep{kinugawa2020formation}. Based on improved binary statistics derived from N-body simulations of Pop~III star clusters \citep{boyuan2020c}, we find that both channels can explain the merger rate of events like GW190521. 

However, agreement with observation in the CE channel can only be (marginally) achieved with no PPI effect, 
or SN kicks to shrink the binary orbits \citep{tanikawa2020merger}, or much smaller Pop~III cluster sizes than expected from small-scale simulations of Pop~III protostar systems, and also requires significant late-time ($z\lesssim 6$) Pop~III star formation. For the NSC DH channel, on the other hand, the observed merger rate is naturally explained with typical populations of NSCs in dark matter halos with $M_{\rm h}\sim 10^{8}-10^{10}\ \rm M_{\odot}$ formed at $z\gtrsim 5.5$. Besides, as pointed out in \citet{gayathri2020gw190521} and \citet{romero2020gw190521}, GW190521 resulted from a system that was either strongly precessing or eccentric, which is unlikely for CE evolution. However, these features are possible for initially wide Pop~III binaries resulting from Pop~III star cluster dynamics which later fall into NSCs, especially when considering the effect on binary orbits by central massive BHs likely co-existing with NSCs (e.g., \citealt{zhang2019gravitational}). Therefore, we emphasize the importance of alternative BHB evolution channels, including environmental effects in addition to the classical binary stellar evolution channel for isolated binaries. Our results based on simplified modelling of the dynamics of Pop~III BHBs in high-$z$ galaxies should be regarded as broad estimations for the range of possible outcomes. We call for more advanced theoretical models to reduce the uncertainties, and further take into account halo growth and mergers, gas accretion and central massive BHs.

Although our analysis favours the NSC DH channel, the CE channel cannot be ruled out, due to significant uncertainties in the two channels. Both may have non-negligible contributions to the BHB mergers in the PPISN mass gap. Moreover, as the former has typical long delay times of a few Gyr, while the latter is dominated by short delay times of a few Myr. Therefore, most of the corresponding BH mergers happen at $z\lesssim 2$ and $z\sim 10$, respectively. With the third generation of gravitational wave detectors such as the Einstein telescope \citep{punturo2010einstein}, we will be able to measure the merger rate density of BHBs in the PPISN mass gap up to $z\sim 10$. If such sources are dominated by Pop~III progenitors, this will enable us to evaluate the relative importance of the two channels of Pop~III BHB mergers, providing a novel probe of early cosmic structure formation.


\section*{Acknowledgments}
The authors acknowledge the Texas Advanced Computing Center (TACC) for providing HPC resources under XSEDE allocation TG-AST120024.


\bibliographystyle{aasjournal}

\appendix

\section{Justification of the DH models for high-$z$ NSCs}

\label{a1}

\begin{figure*}[htbp]
    \centering
    \includegraphics[width=1\columnwidth]{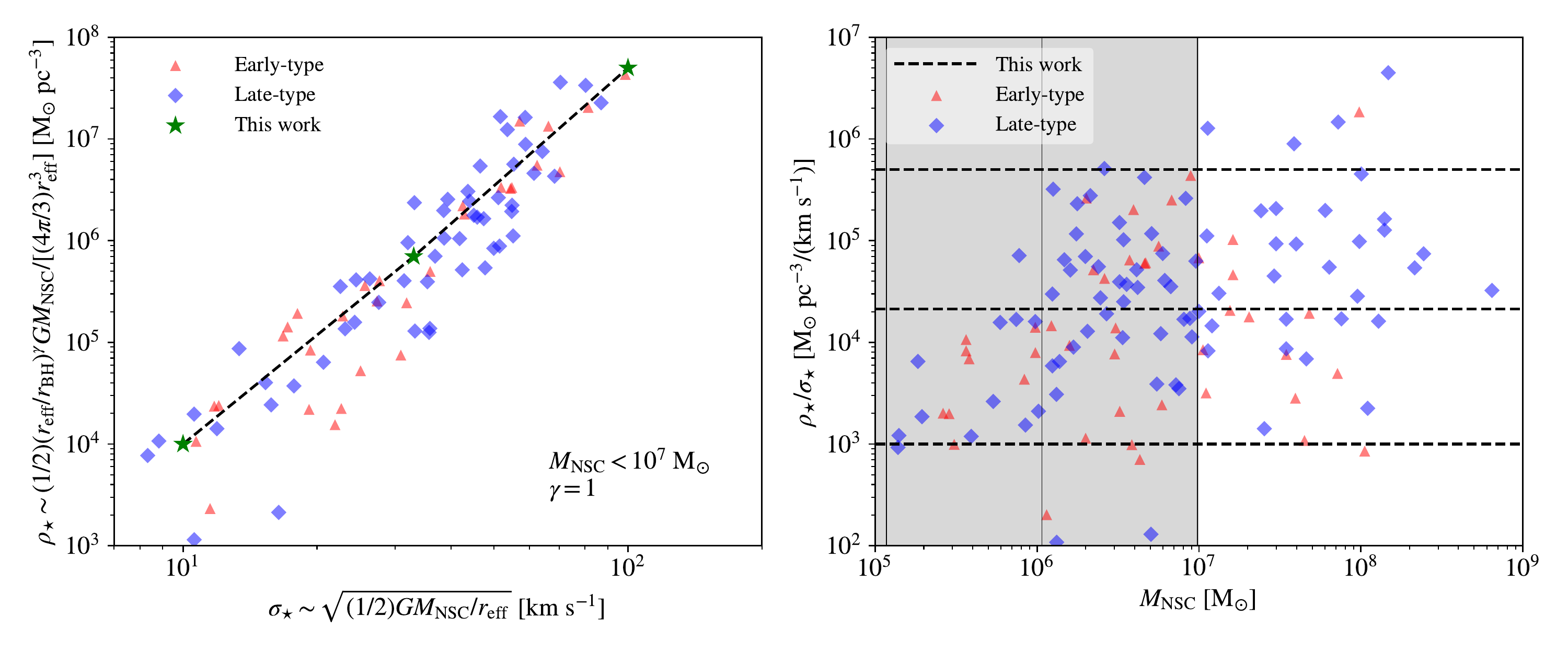}
    \caption{Distributions of observed NSCs from \citet{neumayer2020nuclear} in the $\rho_{\star}$-$\sigma_{\star}$ (left) and $\rho_{\star}/\sigma_{\star}$-$M_{\rm NSC}$ space (right). NSCs in early-type and late-type galaxies are shown with triangles and diamonds, respectively. In the left panel, only NSCs with $M_{\rm NSC}<10^{7}\ \rm M_{\odot}$ are shown. The dashed line denotes the power-law fit to the data, while the stars correspond to the three models adopted in this work for halos with $M_{\rm h}\sim 10^{8},\ 10^{10}\text{ and } 10^{12}\ \rm M_{\odot}$. $r_{\rm BH}$ is the characteristic radius within which the enclosed stellar mass is twice the mass of the infalling BHB ($m\sim 140\ \rm M_{\odot}$). $\gamma=1$ is the inner slope of the stellar density profile \citep{dehnen1993family}. In the right panel, the dashed horizontal lines correspond to the values of $\rho_{\star}/\sigma_{\star}$ in our three models. The thin solid vertical lines show the NSC masses in halos with $M_{\rm h}\sim 10^{8},\ 10^{10}\text{ and } 10^{12}\ \rm M_{\odot}$, where we have used the NSC-stellar mass scaling relation (equ.~1 of \citealt{neumayer2020nuclear}) and assumed that the total stellar mass follows $M_{\star}\sim 0.01M_{\rm h}$.}
    \label{fa1}
\end{figure*}

In this section, we further justify our choices of parameters in Sec.~\ref{s2.3} for NSC properties and dynamics of Pop~III BHs/BHBs in high-$z$ galaxies. The NSC properties adopted in our three models are based on the mass-size scaling relation for NSCs observed in the local Universe (see fig.~7 of \citealt{neumayer2020nuclear}). We assume that high-$z$ NSCs have similar properties as the relaxation timescales in NSCs are usually large (a few Gyr). The distributions of observed NSCs in the $\rho_{\star}$-$\sigma_{\star}$ and $\rho_{\star}/\sigma_{\star}$-$M_{\rm NSC}$ space are shown in Fig.~\ref{fa1}. It turns out that our models capture typical halos with $M_{\rm h}\sim 10^{8},\ 10^{10}\text{ and } 10^{12}\ \rm M_{\odot}$, which cover 93\% of the observed NSC sample in the distribution\footnote{According to Equations~\ref{e1}-\ref{e4}, $\rho_{\star}/\sigma_{\star}$ is the one single parameter that reflects NSC properties in the DH scheme.} of $\rho_{\star}/\sigma_{\star}$. Here we estimate $\sigma_{\star}$ with the circular velocity at the half-mass(light) radius. As Pop~III BHs/BHBs are much more massive than typical stars in the NSC, they will quickly segregate into the cluster center, so that we need to set $\rho_{\star}$ to the central density. However, central volume densities of NSCs are only measured in a few nearby galaxies due to limitation of resolution. To capture the typical stellar density felt by the Pop~III BHB in the cluster center, we set $\rho_{\star}$ to the stellar density within a characteristic radius $r_{\rm BH}$ where the enclosed stellar mass is twice the mass of the infalling BHB ($m\sim 140\ \rm M_{\odot}$). In our case, $r_{\rm BH}\sim 0.002-0.2$~pc, with a median of 0.02~pc. The values of $\rho_{\star}$ obtained in this way are consistent with observation (i.e., $\sim 10^{6}-10^{7}\ \rm M_{\odot}\ pc^{-3}$ at $r_{\rm BH}\sim 0.01-0.1$~pc, \citealt{neumayer2020nuclear}).

\begin{figure}[htbp]
    \centering
    \includegraphics[width=0.5\columnwidth]{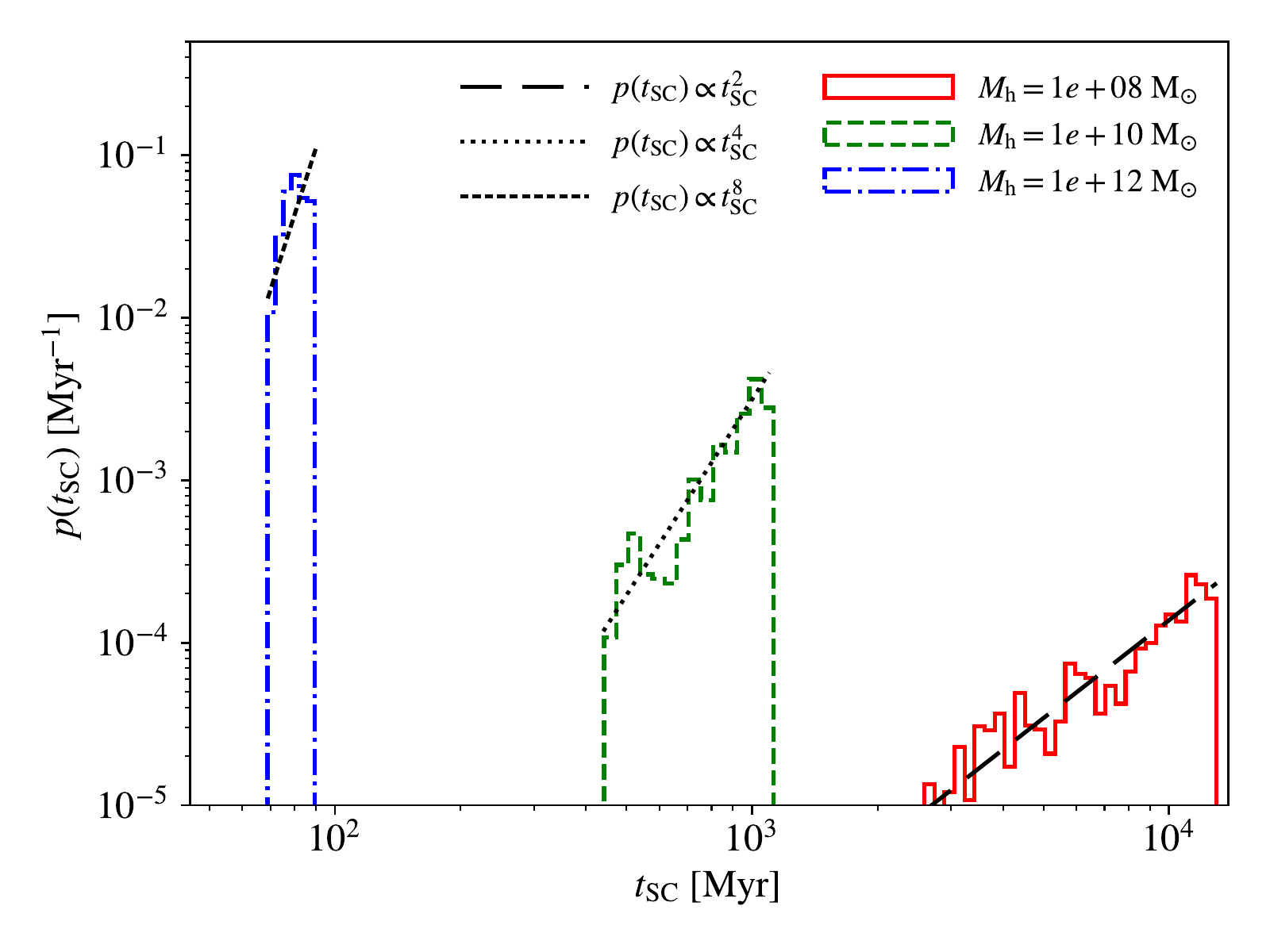}
    \caption{Distributions of the inspiral time $t_{\rm SC}$ in NSCs based on the binary statistics from the FD model in \citet{boyuan2020c} with PPI effect, for the three models with $M_{\rm h}\sim 10^{8},\ 10^{10}\text{ and } 10^{12}\ \rm M_{\odot}$, denoted by the solid, dashed and dashed-dotted contours, respectively. Power-law fits to the distributions are also shown, which are used in the calculation of $\mathcal{P}(t_{\rm GW})$. Note that only \textit{hard} binaries are considered here, which will be further hardened (instead of destroyed) by 3-body interactions. These binaries must satisfy $Gm_{1}m_{2}(1-e)/[a(1+e)]>m_{\star}\sigma_{\star}^{2}$ in order to survive disruptions close to the apocenter, where $m_{\star}=1\ \rm M_{\odot}$ is the typical mass of surrounding stars. }
    \label{fa3}
\end{figure}

Fig.~\ref{fa3} shows the inspiral time distribution $p(t_{\rm SC})$ for the three models with $M_{\rm h}\sim 10^{8},\ 10^{10}\text{ and } 10^{12}\ \rm M_{\odot}$. These distributions can be well-approximated with power-laws of positive slopes $\alpha=2\sim8$, which are quite different from the inspiral time distributions for the CE channel with negative slopes $\alpha=-1\sim-3$ (e.g., \citealt{belczynski2017likelihood,tanikawa2020merger}). The reason is that for the CE channel, the negative slope is associated with the distribution of initial separation $a_{0}$, while in the DH channel, the inspiral time $t_{\rm SC}$ is insensitive to $a_{0}$. As shown in Equ.~\ref{e3}, the term involving $a_{0}$ is negligible for wide binaries ($a_{\rm \star/GW}$ is typically a few AU, much smaller than $a_{0}$). Therefore, $p(t_{\rm SC})$ by DH is determined by the distributions of eccentricity and mass. In general, higher eccentricity and higher total mass lead to smaller $t_{\rm SC}$. For the FD model, the distributions of total mass and eccentricity are both (quasi-)uniform (see fig. 9 in \citealt{boyuan2020c}). In this way, binaries with small $t_{\rm SC}$ must be massive and meanwhile highly eccentric, which are rare, such that $p(t_{\rm SC})$ is dominated by binaries with large $t_{\rm SC}$, and therefore have positive slopes.

\begin{figure}[htbp]
    \centering
    \includegraphics[width=1\columnwidth]{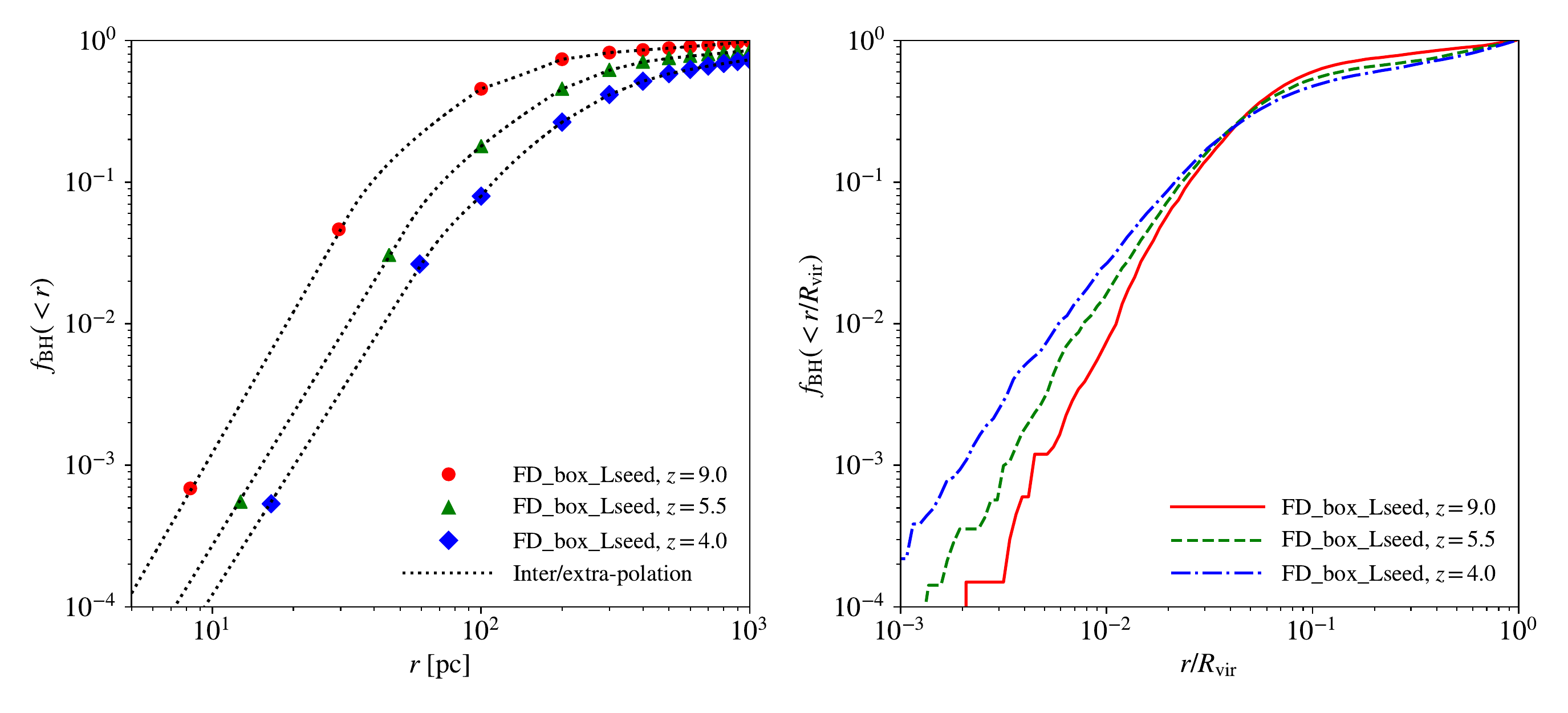}
    \caption{Enclosed fraction of Pop~III BHs in terms of the physical distance $r$ to the galaxy center $r$ (left) and the ratio $r/R_{\rm vir}$ (right), given the halo viral radius $R_{\rm vir}$, for atomic-cooling halos ($M_{\rm h}\gtrsim 10^{8}\ \rm M_{\odot}$, including sub-halos) at $z=9$ (squares and solid), 5.5 (triangles and dashed) and 4 (diamonds and dashed-dotted), from the cosmological simulation \texttt{FD\_box\_Lseed} in \citealt{boyuan2020}. Interpolation and extrapolation (at $r\lesssim 30-60$~pc) of the data are also shown with dotted curves in the left panel, which are used in the calculation of $\mathcal{P}(t_{\rm GW})$.}
    \label{fa2}
\end{figure}

Finally, Fig.~\ref{fa2} shows the enclosed fraction $f_{\rm BH}$ of Pop~III BHs in terms of the physical distance $r$ to the galaxy center and the ratio $r/R_{\rm vir}$, given the halo viral radius $R_{\rm vir}$, for atomic-cooling halos ($M_{\rm h}\gtrsim 10^{8}\ \rm M_{\odot}$, including sub-halos), measured from the cosmological simulation \texttt{FD\_box\_Lseed} in \citet{boyuan2020}, at $z=9$, 5.5 and 4, corresponding to the NSC models with $M_{\rm h}\sim 10^{8},\ 10^{10}\text{ and } 10^{12}\ \rm M_{\odot}$, respectively. In general, at fixed $r$, $f_{\rm BH}(<r)$ increases with redshift. 
The reason is that most Pop~III BHs are formed in low-mass (sub)halos ($M_{\rm h}\sim 10^{7}-10^{8}\ \rm M_{\odot}$) at $z\sim 10$. These low-mass halos then grow or merge into more massive halos at later times, which tend to have larger physical sizes and longer dynamical timescales, such that an increased fraction of Pop~III BHs will not be able to settle into the inner region of $r<1$~kpc. 
Note that dynamical fricition is not resolved at small scales ($\lesssim 100$~pc) in \texttt{FD\_box\_Lseed}, so that the results in Fig.~\ref{fa2} can be regarded as initial distributions of BHs in which the effect of dynamical friction has not kicked in.

\label{lastpage}
\end{document}